\documentclass[conference]{IEEEtran}
\IEEEoverridecommandlockouts
\usepackage{cite}
\usepackage[caption=false]{subfig}
\usepackage{amsmath,amssymb,amsfonts}
\usepackage{algorithmic}
\usepackage{graphicx}
\usepackage{textcomp}
\usepackage{xcolor}
\usepackage{multirow}
\usepackage{balance}
\usepackage{hyperref}

\def\BibTeX{{\rm B\kern-.05em{\sc i\kern-.025em b}\kern-.08em
    T\kern-.1667em\lower.7ex\hbox{E}\kern-.125emX}}
\begin{document}

\title{Adaptive Central Frequencies Locally Competitive Algorithm for Speech\\
\thanks{This research is funded by ''Fonds de recherche du Québec - Nature et technologies'' and the ''Natural Sciences and Engineering Research Council of Canada''.}
}

\author{Soufiyan Bahadi \qquad Eric Plourde \qquad Jean Rouat \\ \textit{NECOTIS Research Lab, Université de Sherbrooke}, Sherbrooke (QC), Canada \\
\{soufiyan.bahadi, eric.plourde, jean.rouat\}@usherbrooke.ca}

\maketitle

\begin{abstract}
\end{abstract}
Neuromorphic computing, inspired by nervous systems, revolutionizes information processing with its focus on efficiency and low power consumption. Using sparse coding, this paradigm enhances processing efficiency, which is crucial for edge devices with power constraints. The Locally Competitive Algorithm (LCA), adapted for audio with Gammatone and Gammachirp filter banks, provides an efficient sparse coding method for neuromorphic speech processing. Adaptive LCA (ALCA) further refines this method by dynamically adjusting modulation parameters, thereby improving reconstruction quality and sparsity. This paper introduces an enhanced ALCA version, the ALCA Central Frequency (ALCA-CF), which dynamically adapts both modulation parameters and central frequencies, optimizing the speech representation. Evaluations show that this approach improves reconstruction quality and sparsity while significantly reducing the power consumption of speech classification, without compromising classification accuracy, particularly on Intel’s Loihi 2 neuromorphic chip.

\section{Introduction}
Neuromorphic computing, inspired by nervous systems \cite{neuromorphic1, neuromorphic2}, is a shift towards efficient and low-power information processing \cite{lowp1, lowp2}. It emulates neural computation and leverages sparse coding, reducing computational demands compared to dense representations. This approach enhances the efficiency of brain-inspired AI systems, especially for power-constrained edge devices.

The Locally Competitive Algorithm (LCA) \cite{lca} is an example of an efficient implementation for sparse coding. Due to its neural architecture, LCA is well-suited for implementation on neuromorphic hardware \cite{slcaloihi1, slcaloihi2}, enabling real-time processing while being energy-efficient. In neuromorphic speech processing, LCA has recently demonstrated its potential in tasks such as the classification of spoken digits and command words \cite{preprint}. It has been used to represent raw speech signals with events, offering more accurate speech classification compared to a software-based silicon cochlea model named LAUSCHER \cite{shd}. However, this accuracy improvement comes at the expense of increased power consumption. Adaptive LCA (ALCA) \cite{alca} has been employed in \cite{preprint} to enhance the power efficiency of speech classification systems while maintaining the accuracy achieved with LCA representation. This highlights the potential contribution that LCA and ALCA can bring to neuromorphic speech processing.

This article aims to continue this line of work by exploring further improvements to LCA/ALCA and studying the extent to which the energy efficiency of LCA, as a sparse coding technique for speech processing, can be enhanced. ALCA serves as an example of leveraging LCA’s neural architecture to improve its capabilities. This framework introduces a dynamic adaptation mechanism for the modulation parameters of the Gammachirp filter bank, using the backpropagation algorithm to guide the adaptation, ultimately improving the sparsity and reconstruction quality of the representation.

In this article, we introduce an enhanced version of ALCA, termed ALCA - Central Frequency (ALCA-CF), which further optimizes the algorithm’s efficiency by dynamically adapting not only the modulation parameters of the cochlear filters but also their central frequencies. 

The idea of adapting filter bank frequencies has previously been proposed in SincNet \cite{sincnet}, a learning approach that replaces the first layer of a neural network processing raw audio with a filter bank using sinc functions. This method learns the low and high cutoff frequencies of band-pass filters. This approach addresses the problem in end-to-end methods where the first layer not only handles high-dimensional inputs, but also suffers more from the gradient vanishing problem than the other layers. However, the learned frequency distribution is optimized for the specific task that SincNet is used for and does not inherently focus on enhancing representation sparsity or reducing redundancy.

In contrast, ALCA-CF’s adaptation of central frequencies is grounded in the LCA context, targeting both improved reconstruction quality and increased sparsity. This emphasis on sparsity is particularly advantageous for neuromorphic computing, as it directly impacts the activity of spiking neural networks, thereby effecting power consumption. Unlike SincNet’s task-specific frequency adaptation, ALCA-CF employs a data-driven mechanism, making its representations versatile and potentially applicable across a wide range of audio tasks.

Our results demonstrate that ALCA-CF improves reconstruction quality and sparsity compared to ALCA. Crucially, ALCA-CF reduces power consumption in speech classification, particularly when implemented on Intel’s Loihi 2 neuromorphic chip \cite{loihi2}. This establishes ALCA-CF as a robust alternative for efficient and adaptive audio representation in neuromorphic systems.

\section{Methods}
\subsection{Locally Competitive Algorithm (LCA)}
\label{ssec:lca}

The goal of LCA \cite{lca} is to represent an input signal $\boldsymbol{s}$ as a linear combination of a family of vectors (atoms) $\boldsymbol{D} = [\boldsymbol{\phi}_1, ..., \boldsymbol{\phi}_N]$, called a ``dictionary'', where most coefficients $\boldsymbol{a} = [a_1, ..., a_N]^T$ are zero:
\begin{equation}
    \label{eq:recons}
    \displaystyle
    \boldsymbol{\hat{s}} = \sum_{i=1}^{N} a_{m} \boldsymbol{\phi_m} = \boldsymbol{D}\boldsymbol{a},
\end{equation}
where $\boldsymbol{\hat{s}}$ is the approximation of the input signal $\boldsymbol{s}$ and $N$ is the number of atoms. To achieve this, a recurrent neural network incorporating lateral inhibition is defined with an objective function to be minimized. This function is referred to as an energy function $E$. It is defined as a combination of the Mean Squared Error (MSE) between $\boldsymbol{s}$ and $\boldsymbol{\hat{s}}$, a sparsity cost penalty $S$ evaluated from the activation of neurons that corresponds to the coefficients $\boldsymbol{a}$ in (\ref{eq:recons}), and a Lagrange multiplier $\lambda$.
\begin{equation}
    \label{eq:energy}
    \displaystyle
    E = \frac{1}{2}||\boldsymbol{\hat{s}}-\boldsymbol{s}||^2 + \lambda S(\boldsymbol{a}).
\end{equation}
The minimization of this energy function is carried by the neural dynamics which are governed by the vectorized ordinary differential equation:
\begin{equation}
    \label{eq:ODE}
    \displaystyle
    \tau \frac{d\boldsymbol{v}}{dt} = \boldsymbol{p} - \boldsymbol{v} - (\boldsymbol{D}^T\boldsymbol{D}-\boldsymbol{I})\boldsymbol{a},
\end{equation}
where $\tau$ is the time constant of each neuron, $\boldsymbol{p}$ is the input signal projection on the dictionary, i.e., $\boldsymbol{p} = \boldsymbol{D}^T \boldsymbol{s}$, $\boldsymbol{v}$ is the membrane potential vector, and $\boldsymbol{I}$ is the identity matrix. Mainly, the evolution of $\boldsymbol{v}$ over time depends on the input intensity $\boldsymbol{p}$ as an integration term and on $-\boldsymbol{v}$ as a leak term which makes these neurons behave like leaky integrators. Membrane potentials exceeding the threshold $\lambda$ –––the same as the Lagrange multiplier in (\ref{eq:energy})––– produce activations corresponding to the coefficients $\boldsymbol{a}$. Each activated neuron inhibits all others through horizontal connections, whose weights are calculated as $\boldsymbol{D}^T\boldsymbol{D}-\boldsymbol{I}$. For each neuron $m$, the activation $a_m$ is a non-linearity $T_\lambda$ that can be sigmoidal or –––as we used in this work––– the hard thresholding function applied to the potential $v_m$ of neuron $m$:

\begin{equation}
    \label{eq:thresh}
    \displaystyle
    a_m = T_\lambda(v_m) =
    \begin{cases}
      0 & \text{if $|v_m| < \lambda$}\\
      v_m & \text{otherwise}
    \end{cases}.
\end{equation}
As shown in \cite{lca}, $\frac{d\boldsymbol{v}}{dt} \propto -\frac{\partial E}{\partial \boldsymbol{a}}$, meaning the energy function (\ref{eq:energy}) is minimized.

\subsection{Dictionary}
\label{sec:dict}
For audio applications \cite{plca, clca}, the dictionary $\boldsymbol{D}$ (\ref{eq:recons}) comprises Gammachirp filters impulse responses \cite{Gammachirp}, which are a frequency-modulated carrier (a chirp) with an envelope that follows a Gamma distribution function. for a channel $i$:
\begin{equation}
    \label{eq:gam}
    \displaystyle
    g_i(t) = t^{l_i-1} e^{-2 \pi b_i \text{ERB}(f_i)t} \cos(2 \pi f_i t + c_i \ln(t)),
\end{equation}
where $l_i$ and $b_i$ are Gamma distribution parameters that control the attack and the decay of the filter, $c_i$ is referred to as the chirp parameter which modulates the carrier frequency allowing to slightly modify the instantaneous frequency, and $f_i$ is the central frequency. $\text{ERB}(f_i)$ is a linear transformation of $f_i$ on the Equivalent Rectangular Bandwidth scale \cite{glasberg}:

We create with (\ref{eq:gam}) a discrete-time dictionary by striding each sampled filter of size $F_l$ across the sampled signal with a stride $r$.

\subsection{Adaptive Locally Competitive Algorithm (ALCA)}
\label{sec:alca}
As shown in \cite{alca}, there is a differentiable relationship between the energy function (\ref{eq:energy}) and the modulation parameters of the Gammachirp. The gradient of the energy function with respect to the modulation parameters is:
\begin{align}
    \displaystyle
    \begin{aligned}
        \frac{\partial E}{\partial (\boldsymbol{c}|\boldsymbol{b}|\boldsymbol{l})} = & \frac{\partial \frac{1}{2}||\boldsymbol{D}\boldsymbol{a} - \boldsymbol{s}||^2}{\partial \boldsymbol{D}\boldsymbol{a}} \frac{\partial \boldsymbol{D}\boldsymbol{a}}{\partial \boldsymbol{D}} \frac{\partial \boldsymbol{D}}{\partial (\boldsymbol{c}|\boldsymbol{b}|\boldsymbol{l})}\\ & + \alpha \lambda \frac{d S(\boldsymbol{a})}{d \boldsymbol{a}} \frac{d \boldsymbol{a}}{d \boldsymbol{v}} \frac{\partial \boldsymbol{v}}{\partial \boldsymbol{D}} \frac{\partial \boldsymbol{D}}{\partial (\boldsymbol{c}|\boldsymbol{b}|\boldsymbol{l})}
    \end{aligned},
    \label{eq:chain}
\end{align}
where $|$ is the "OR" operator and $\boldsymbol{c}$, $\boldsymbol{b}$, and $\boldsymbol{l}$ are vectors containing the parameters of the filters $c_i$, $b_i$, and $l_i$, respectively. A trade-off parameter $\alpha$ is introduced to decouple LCA's neurons threshold $\lambda$ in (\ref{eq:thresh}) from the Lagrange multiplier $\lambda$ in (\ref{eq:energy}). This relationship enables the implementation of the gradient descent, allowing the gradient flow from the membrane potential to backpropagate further towards the Gammachirp's parameters \cite{alca}. 

\subsection{Adaptation of Central Frequency Distribution with ALCA (ALCA-CF)}
In this work, we propose incorporating the distribution of central frequencies into the adaptation mechanism of ALCA. The Gammachirp function (\ref{eq:gam}) is differentiable with respect to its central frequency $f_i$. Therefore, the gradient of the energy function (\ref{eq:energy}) can backpropagate towards $f_i$:

\begin{equation*}
    \label{eq:chain2}
    \displaystyle
        \frac{\partial E}{\partial \boldsymbol{f}} = \frac{\partial \frac{1}{2}||\boldsymbol{D}\boldsymbol{a} - \boldsymbol{s}||^2}{\partial \boldsymbol{D}\boldsymbol{a}} \frac{\partial \boldsymbol{D}\boldsymbol{a}}{\partial \boldsymbol{D}} \frac{\partial \boldsymbol{D}}{\partial \boldsymbol{f}} + \alpha \lambda \frac{d S(\boldsymbol{a})}{d \boldsymbol{a}} \frac{d \boldsymbol{a}}{d \boldsymbol{v}} \frac{\partial \boldsymbol{v}}{\partial \boldsymbol{D}} \frac{\partial \boldsymbol{D}}{\partial \boldsymbol{f}}
\end{equation*}
where $\boldsymbol{f}$ is the vector made of central frequencies $f_i$. This method allows adapting the distribution of central frequencies that minimizes (\ref{eq:energy}).

The proposed adaptation of Central Frequencies with ALCA (ALCA-CF) starts from a dictionary comprising Gammatones (GT) atoms, which are a specific type of Gammachirp as defined in \cite{glasberg}, having their frequencies in the log-scale. We then use the truncated backpropagation through time (TBPTT) \cite{TBPTT} and Adamax optimizer \cite{adam} to adapt the central frequencies along with modulation parameters.

The advantage of ALCA-CF lies in its ability to provide greater flexibility for filters to effectively compete in representing the input signal. Whereas ALCA restricts filters to adjust sensitivity around a fixed central frequency, ALCA-CF extends this capability by enabling filters to dynamically shift along the frequency scale. This enables the formation of a non-linear frequency resolution, which adapts to the acoustical characteristics of the input. Crucially, this adaptation process is guided by the energy function (\ref{eq:energy}), ensuring that the representation remains sparse.

\subsection{Datasets and Representations}
\label{sec:dataset}
This study focuses on speech processing using two datasets. The first is Heidelberg Digits (HD) \cite{shd}, which contains approximately $10\thinspace{}420$ recordings of spoken digits ($0$–$9$) in English and German from $12$ speakers, sampled at $48$ kHz. The dataset is divided into training and test sets. Two speakers exclusively reserved for testing, and $5\%$ of test samples drawn from speakers in the training set to evaluate speaker generalization. The second is Google Speech Commands (SC) \cite{sc}, consisting of $105\thinspace{}829$ one-second audio files sampled at $16$ kHz, representing $35$ classes of English words spoken by $1\thinspace{}864$ speakers. We use version $2$ of this dataset, partitioned into training, testing, and validation sets using the hashing function from \cite{shd}.

The HD dataset has a spiking version, referred to as spiking HD (SHD), based on LAUSCHER's artiﬁcial cochlea model \cite{shd} that converts audio speech data to a spike train representation with 700 channels, similar to a spectrogram representation with Mel-spaced ﬁlter banks. We derive from this representation the temporal histograms of SHD obtained by time binning the analog spike times in time bins of 10 ms as in \cite{shd}.

We also encode the audio recordings of HD and SC using LCA, selecting $k = 700$ filters to match the number of channels in SHD. To maintain a similar analysis window of $10$ ms, as used in the temporal histograms of SHD, we set the filter length ($F_l$) and the stride ($r$) to $F_l = 1024$ and stride $r=512$ for HD, and $F_l = 256$ and $r=128$ for SC. To ensure comparable sparsity with the LAUSCHER representation for these datasets, we set the threshold ($\lambda$) to $\lambda = 0.00045$ for HD and $\lambda = 0.0007$ for SC. From this LCA representation, we derive the ALCA representation by adapting the Gammachirps, both excluding and including the central frequencies, which results in two additional representations: ALCA and ALCA-CF.

For ALCA, the adaptation hyper-parameters were optimized in \cite{preprint}. In this article, we use the selected values from \cite{preprint}. We perform the same hyper-parameter optimization for ALCA-CF but we characterize the central frequencies with their own backpropagation learning rate \textit{lr-cf} since their range is from 0 to the Nyquist frequency which is significantly different from the ranges of modulation parameters. The search interval of \textit{lr-cf} is $[10^{-6}, 10^2]$ on a logarithmic scale.

In summary we consider LCA, ALCA, and ALCA-CF for benchmarks on sparsity and reconstruction quality, while SHD is considered for power consumption benchmarks.
\begin{figure*}[!t]
    \centering
    \subfloat[Central frequencies\label{fig:fig1}]{
        \includegraphics[width=0.23\textwidth]{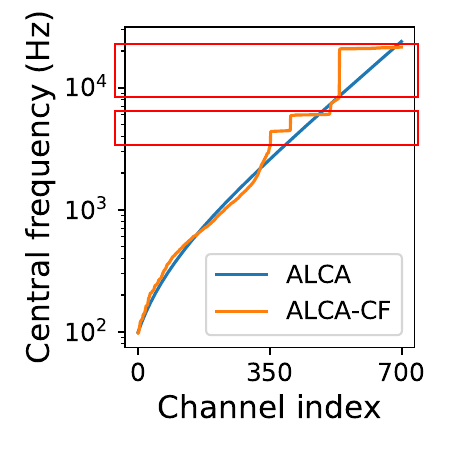}
    }
    \hfill
    \subfloat[LCA representation\label{fig:fig2}]{
        \includegraphics[width=0.23\textwidth]{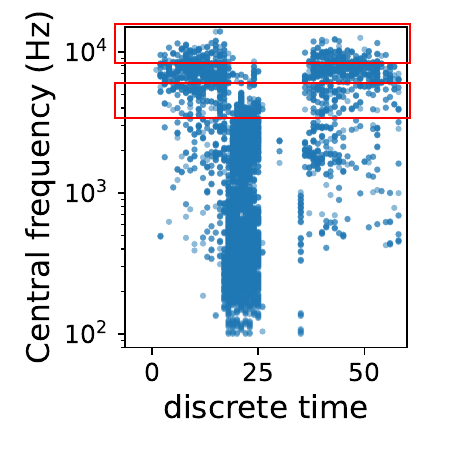}
    }
    \hfill
    \subfloat[ALCA representation\label{fig:fig3}]{
        \includegraphics[width=0.23\textwidth]{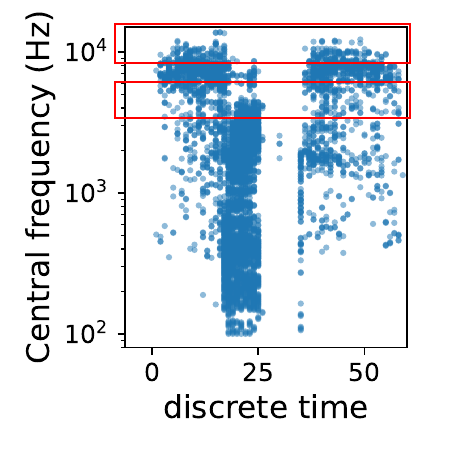}
    }
    \hfill
    \subfloat[ALCA-CF representation\label{fig:fig4}]{
        \includegraphics[width=0.23\textwidth]{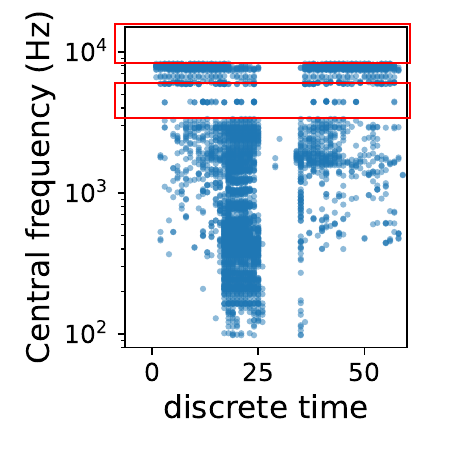}
    }
    
    \caption{The central frequencies used with ALCA and ALCA-CF and the representations of a pronounced "six" signal. The red outlines indicate the intervals where the resolution of central frequencies of ALCA-CF is extremely high.}
    \label{fig:cf}
\end{figure*}

\subsection{Classification Task}
\label{sec:classification}
To evaluate the impact of ALCA-CF's adaptation on performance and power consumption in speech classification, we use the same experimental setup as in \cite{preprint}, employing a single recurrent layer SNN with the current-based Leaky Integrate-and-Fire model and a linear readout layer. The cross-entropy loss is applied to the average output of each readout unit, computed across all time steps. The training is performed using truncated backpropagation through time (TBPTT) \cite{TBPTT} with the fast sigmoid surrogate gradient \cite{superspike}. We use the Adamax optimizer with $L_2$ weight regularization. For more details on weight initialization and the hyper-parameter optimization strategy, please refer to \cite{preprint}. We trained the model with the Lava Deep Learning library (\texttt{Lava-DL}) \cite{lava-dl}.

\section{Results and Discussion}
\label{sec:results}
We first compute the average sparsity and reconstruction quality for each representation in the test set with LCA, ALCA, and ALCA-CF. Reconstruction quality is measured with the Signal-to-Noise Ratio (SNR), where the noise is defined as the difference between the original signal and the reconstructed signal. These results are presented in Tab.~\ref{tab:snr_sp}, showing that for HD, ALCA-CF achieves the highest SNR, exceeding ALCA by $2$ dB and LCA by $5.52$ dB. Furthermore, ALCA-CF provides the sparsest representation, being $11.47\%$ sparser than ALCA and $17.53\%$ sparser than LCA. These results are confirmed with SC, where the disparity between LCA, ALCA, and ALCA-CF is more pronounced. This method enables ALCA-CF to select frequencies that represent audio signals more efficiently compared to the logarithmic scale distribution used with LCA and ALCA.

\begin{table}[t]
    \caption{The average number of active neurons (sparsity) and the average signal-to-noise ratio (SNR) on the test sets of HD and SC with LCA, ALCA, and ALCA-CF.}
    \begin{center}
    \begin{tabular}{ |c|c|c|c|c| }
     \hline
      \textbf{Dataset} & \textbf{Metric} & \textbf{LCA} & \textbf{ALCA} & \textbf{ALCA-CF} \\ 
     \hline
     \multirow{2}{*}{\textbf{HD}} & \textbf{SNR (dB)} & $9.83$ & $13.36$ & \boldmath $15.35$ \\
     \cline{2-5}
     & \textbf{Sparsity} & $7901.91$ & $7360.60$ & \boldmath $6516.33$ \\
     \hline
     \multirow{2}{*}{\textbf{SC}} & \textbf{SNR (dB)} & $13.86$ & $20.51$ & \boldmath $23.04$ \\
     \cline{2-5}
     & \textbf{Sparsity} & $8448.55$ & $7904.70$ & \boldmath $5838.24$ \\
     \hline
    \end{tabular}
    \end{center}
    \label{tab:snr_sp}
\end{table}

To better understand the enhanced selectivity of ALCA-CF compared to ALCA, we present in Fig.\ref{fig:cf} the central frequencies used with ALCA and the adapted ones used with ALCA-CF, along with the representations of a pronounced ''six'' signal. Fig\ref{fig:cf}.a shows that ALCA-CF has modified the frequency resolution. Given that the dataset consists of speech, ALCA-CF is expected to select low frequencies with high resolution and high frequencies with lower resolution. However, ALCA-CF decreases the resolution below $600$ Hz, slightly increases it between $600$ Hz and $2100$ Hz, and then decreases it again below approximately $3500$ Hz. For frequencies above $3500$ Hz, ALCA-CF significantly increases the resolution at various intervals, creating a step-like pattern in the high-frequency distribution. This can be explained by the fact that ALCA-CF requires fewer coefficients in high frequencies to approximate a speech signal compared to low frequencies. Therefore, ALCA-CF brings channels of high central frequencies closer together, making them more similar and thereby increasing their correlation. This results in a higher level of lateral inhibition, represented by $\boldsymbol{D}^T\boldsymbol{D}-\boldsymbol{I}$ in (\ref{eq:ODE}). Consequently, the competition between channels of high frequencies is particularly intense in the ``step'' frequency intervals (outlined regions in Fig.\ref{fig:cf}). 

Furthermore, to enhance the reconstruction quality of speech, ALCA-CF needs more coefficients to represent low frequencies than high frequencies, which is why the frequency resolution for low frequencies does not exhibit the ``step'' effect. This is confirmed by an example of a pronounced "six" represented by LCA, ALCA, and ALCA-CF in Fig.~\ref{fig:cf}.b-d, respectively. Notably, ALCA-CF produces almost no coefficients for high frequencies above $8500$ Hz, unlike LCA and ALCA. For frequencies between $3500$ Hz and $6000$ Hz, there is a significant reduction in the number of coefficients produced by ALCA-CF compared to those produced by ALCA and LCA. Importantly, ALCA-CF reduces the number of coefficients not only in high frequencies but also in low frequencies. Specifically, for frequencies below $3500$ Hz, ALCA-CF produces $2665$ coefficients, while ALCA produces $3041$, and LCA produces $3037$. The overall number of coefficients is reduced by ALCA-CF by more than $12\%$ compared to ALCA and LCA, which produced nearly the same sparsity.

\begin{table}[t]
    \centering
    \caption{Comparison of Dynamic Power Consumption per Inference of Quantized 1024-Neurons Recurrent SNN Alongside the Test Accuracy for SHD, LCA, ALCA, and ALCA-CF.}
    \begin{tabular}{ |c|c|c|c|c| }
     \hline
       \textbf{Representation} & \textbf{SHD} & \textbf{LCA} & \textbf{ALCA} & \textbf{ALCA-CF} \\ 
        & \cite{preprint} & \cite{preprint} & \cite{preprint} & \textbf{(this work)} \\
     \hline
     \textbf{Test Accuracy (\%)} & $83.77$ & $94.63$ & $94.38$ & \boldmath $94.88$ \\
    & $\pm 1.2$ & $\pm 0.8$ & $\pm 1.0$ & \boldmath $\pm 0.9$\\
     \hline
     \textbf{Dynamic Power (W)} & $0.015$ & $0.021$ & $0.013$ & \boldmath $0.004$ \\
     \hline
    \end{tabular}
    \label{tab:bench}
\end{table}

Moving to the classification task, we conduct a performance/power efficiency benchmark on Loihi 2 using SNNs with SHD, LCA, ALCA, and ALCA-CF representations. In this comparison, two metrics are considered: test classification accuracy and dynamic power. Dynamic power refers to the power consumed by the hardware when the system is actively running the workload. Since our main goal is to enhance the sparsity of speech representation and the resulting activity sparsity in the SNN, we consider the dynamic power per inference. Analyzing static power is beyond the scope of this paper, as it is associated with the hardware itself rather than the neural network operating on it. Loihi’s efficiency is measured using the power probes of \texttt{Lava 0.9.0} on \texttt{Oheogultch} board \texttt{ncl-ext-og-05}.

Tab.~\ref{tab:bench} shows that ALCA-CF achieves the highest test accuracy of $94.88\%$, slightly outperforming LCA and ALCA with accuracies of $94.63\%$ and $94.38\%$, respectively, while SHD recorded a lower accuracy of $83.77\%$. Importantly, ALCA-CF excels in power efficiency, consuming just $0.004$ W per inference, lower than LCA ($0.021$ W), ALCA ($0.013$ W), and SHD ($0.015$ W). These results highlight that the classification-to-power performance balance of ALCA is enhanced with ALCA-CF, achieving the most efficient power consumption. Furthermore, data-driven central frequency adaptation does not negatively impact the classification accuracy.


\section{CONCLUSION}
\label{sec:conclusion}

We propose ALCA-CF, an adaptive approach that offers ALCA more flexibility to enhance reconstruction quality and sparsity. It enables the gradient flow from the ALCA's energy function to reach the central frequencies of the dictionary. Therefore, ALCA-CF modifies the resolution of the central frequency distribution such that the reconstruction quality increases with a sparser representation. This sparsity has an impact on the efficiency of the processing SNN. In the context of spoken digits classification, ALCA-CF demonstrates improved power consumption on Loihi2 without compromising the classification accuracy. This makes ALCA-CF a better candidate than ALCA, LCA, and LAUSCHER cochlea model for low-power speech classification tasks.
Preliminary experiments conducted on noisy speech seem to show that ALCA-CF has also a strong potential. In that context, a more detailed study is left for future work.

\section*{ACKNOWLEDGMENTS}
\label{sec:ack}
Thanks to the reviewers for their constructive feedback, NVIDIA for donating the GTX1080 and Titan XP, INTEL for access to Loihi2, and Andreas Wild for insights on power/energy benchmarks.

\balance
\bibliographystyle{IEEEtran}
\bibliography{refs}
\end{document}